\input{epsf}
\documentclass[11pt]{article}
\def\be{\begin{equation}}
\def\eea{\end{eqnarray}}
\def\bea{\begin{eqnarray}}
\def\ee{\end{equation}}

\author{M.Mohammadi$^{}$ \footnote{majid471702@yahoo.com}
\\$^{}$ {\small Department of Physics, Islamic Azad University - Shahreza Branch, Shahreza, Isfahan, Iran}
}
\title{The effective mass of atom-radiation field system and the cavity-field Wigner distribution
 in the presence of a homogeneous gravitational field
 in the Jaynes-Cummings
           model}
 \begin{document}
\maketitle
\begin{abstract}
\noindent The effective mass that approximately describes the
effect of a classical homogeneous gravitational field on an
interacting atom-radiation field system is determined within the
framework of the Jaynes-Cummings model.
 By taking into account both the atomic motion and gravitational field, a full quantum treatment
 of the internal and external dynamics of the atom is presented.
  By solving exactly the Schr\"{o}dinger equation in the interaction picture, the evolving state
   of the system is found. Influence of a classical homogeneous gravitational field
   on the energy eigenvalues, the effective mass of atom-radiation field system and the Wigner distribution of
the radiation field are studied, when initially
  the radiation field is prepared in a coherent state and the two-level atom is
in a coherent superposition of the excited and ground states.
\end{abstract}
\\
\section{Introduction}
There are several theoretical schemes for the problem of
describing the action of cavity in quantum optics [1-4]. The
effective mass approach is one of these theoretical schemes.
Larson et al. investigate the effect of
cavity on the atomic motion [5]. They introduced the
effective mass parameter by using the Floquet theory [6-8]. In
this theory, the photonic wavelength is small compared with the
cavity length and the system can be treated approximately as
periodic. The spectra of periodic Hamiltonians are known to
consist of allowed energies in forms of bands, separated by
forbidden gaps. These articles, which deal with the action of
cavity in quantum optics, are based on the Jaynes-Cummings model
(JCM) [9]. The JCM of a two-level atom interacting with a single
quantized mode of the electromagnetic field in a lossless cavity
within the rotating wave approximation (RWA) is one of the few
exactly solvable quantum mechanical models in quantum optics. In
the framework of this model many nonclassical effects of
atom-radiation field interactions, such as Rabi oscillations
[10,11], collapse-revival phenomena [12-14], sub-Poissonian
statistics [15] and squeezing of the radiation field [16] have been predicted.\\
\hspace*{00.5 cm} The recent experiments with Rydberg atoms
[17-19] have allowed the study of the dynamics of the two-level
atom interacting with a single mode of the radiation field and to
test the above mentioned nonclassical effects. The standard JCM,
however, is not always valid. In this model, the atom is either
assumed to be studied still relative to the cavity mode, or to have a
large kinetic energy; in both cases the atomic motion is
described classically and the kinetic-energy term may be excluded
from the Hamiltonian. But, if the kinetic energy of the atom is
of the same order of magnitude as the atom-radiation field
interaction energy, the dynamics is significantly changed [20].
Thus, for every cold atom (ion) [21], the kinetic energy term for
the atomic center
of mass motion must be treated quantum mechanically.\\
\hspace*{00.5 cm}On the other hand, experimentally, atomic beams
with very low velocities are generated in laser cooling and atomic
interferometry [22]. It is obvious that for atoms moving with a
velocity of a few millimeters or centimeters per second for a time
period of several milliseconds or more, the influence of Earth's
acceleration becomes important and cannot be neglected [23]. For
this reason, it is of interest to study the temporal evolution of
a moving atom simultaneously exposed to the gravitational field
and a single-mode cavity-field. Since any quantum optical
experiment in the laboratory is actually made in a non-inertial
frame it is important to estimate the influence of Earth's
acceleration on the outcome of the experiment. By referring to the
equivalence principle, one can get a clear picture of what is
going to happen in the interacting atom-field system exposed to a
classical homogeneous gravitational field [24]. A semi-classical
description of a two-level atom interacting with a running laser
wave in a gravitational field has been studied [25]. However, the
semi-classical treatment does not permit the study of the pure
quantum effects occurring in the course of atom-field interaction.
Within a quantum treatment of the internal and external
dynamics of the atom, a theoretical scheme has recently presented [26]
based on an su(2) algebraic structure to investigate the
influence of a classical homogeneous gravitational field on the
quantum non-demolition measurement of atomic momentum in the
dispersive JCM. Also, the effects of the
gravitational field on quantum statistical properties of the
lossless [24] as well as the phase-damped JCMs [27] have investigated. It has been found
that the gravitational field seriously suppresses non-classical
properties
of both the cavity-field and the moving atom.\\
 \hspace*{00.5 cm} In this paper first the effective mass of atom-radiation field system is obtained and then the
  influence of a classical homogeneous gravitational field on the energy eigenvalues, the effective mass
  of atom-radiation field system
    and the Wigner distribution of
the radiation field of the JCM are investigated. In the Jaynes-Cummings model,
when the atomic motion is in a propagating light wave, a two-level atom interacting with the quantized
cavity-field in the presence of a homogeneous gravitational
field is considered. By solving the Schr\"{o}dinger equation in the interaction
picture, the evolving state of the system is found.
In section 2, a quantum treatment of the
internal and external dynamics of the atom with an alternative
su(2) dynamical algebraic structure within the system is presented. Based on
this su(2) structure and in the interaction picture, first
 the effective Hamiltonian describing the atom-field
interaction in the presence of a classical gravity field is obtained and then
 the effective mass of the system is gained. In section 3, we investigate
  the dynamical evolution of atom-radiation field system and show that
how the gravity may affect the Wigner distribution of the radiation field, when initially
  the radiation field is prepared in a coherent state and the two-level atom is
in a coherent superposition of the excited and ground states. The summarized conclusions are presented in section 4.
\section{ The Effective Mass for the JCM in the Presence of Gravitational Field}
The Hamiltonian of atom-field system
in the presence of gravitational field with the atomic motion
along the position vector $\hat{\vec{x}}$ and in the rotating
wave approximation is given by\\
\\
\begin{eqnarray}
\hat{H}=&&\frac{\hat{p}^{2}}{2M}-M\vec{g}.\hat{\vec{x}}+\hbar\omega_{c}(\hat{a}^{\dag}\hat{a}+\frac{1}{2})+\frac{1}{2}\hbar\omega_{eg}\hat{\sigma}_{z}+\nonumber\\
&&\hbar\lambda[\exp(-i\vec{q}.\hat{\vec{x}})\hat{a}^{\dag}\hat{\sigma}_{-}+\exp(i\vec{q}.\hat{\vec{x}})\hat{\sigma}_{+}\hat{a}],
\end{eqnarray}
where $\hat{a}$ and $\hat{a}^{\dag}$ denote, respectively, the
annihilation and creation operators of a single-mode traveling
wave with frequency $\omega_{c}$, $\vec{q}$ is the wave vector of
the running wave and $\hat{\sigma}_{\pm}$ denote the raising and
lowering operators of the two-level atom with electronic levels
$|e\rangle, |g\rangle $ and Bohr transition frequency
$\omega_{eg}$. The atom-field coupling is given by the parameter
$\lambda$ and
 $\hat{\vec{p}}$, $\hat{\vec{x}}$
denote, respectively, the momentum and position operators of the
atomic center of mass motion and $g$ is Earth's gravitational
acceleration. It has been shown [26] that based on su(2) algebraic
structure, as the dynamical symmetry group of the model, the
Hamiltonian (1) can be transformed to the following effective
Hamiltonian
\begin{equation}
\hat{H}_{eff} =\hbar\omega_{c}\hat{K} -\hbar
\hat{\triangle}(\hat{\vec{p}},\vec{g},t)\hat{S}_{0}+\hbar \lambda
( \sqrt{\hat{K}}\hat{S}_{-}+\sqrt{\hat{K}}\hat{S}_{+}),
\end{equation}
where the operators
\begin{equation}
\hat{S_{0}}=\frac{1}{2}(|e \rangle \langle e|-|g \rangle \langle
g|) , \hat{S_{+}}=\hat{a}|e\rangle \langle
g|\frac{1}{\sqrt{\hat{K}}},
\hat{S_{-}}=\frac{1}{\sqrt{\hat{K}}}|g\rangle \langle
e|\hat{a}^{\dag},
\end{equation}
with the following commutation relations
\begin{equation}
[\hat{S_{0}},\hat{S_{\pm}}]=\pm
\hat{S_{\pm}},[\hat{S_{-}},\hat{S_{+}}]=-2\hat{S_{0}},
\end{equation}
are the generators of the su(2) algebra and the operator
\begin{equation}
\hat{\triangle}(\hat{\vec{p}},\vec{g},t)=\frac{3}{2}
(\omega_{c}-(\omega_{eg}+\frac{2\vec{q}.\hat{\vec{p}}}{3M}+\frac{2\vec{q}.\vec{g}t}{3})),
\end{equation}
has been introduced as the Doppler shift detuning at time $t$
[26]. The Hamiltonian (2) has the form of the Hamiltonian of the
JCM, the only modification being the dependence of the detuning
on the conjugation momentum and the gravitational field. The
eigenvalues and eigenstates of the Hamiltonian (2) are
respectively given by [28]
\begin{equation}
E_{\pm,n}(\vec{p},\vec{g},t)=\hbar\omega_{c}(n+1)\pm
\hbar\sqrt{\triangle(\vec{p},\vec{g},t)^{2}+4\lambda^{2}(n+1)},
\end{equation}
\begin{equation}
|+,n,\vec{p}\rangle=\cos\vartheta_{n}(\vec{p},\vec{g},t)|e\rangle\otimes|n\rangle \otimes |\vec{p}\rangle
+\sin\vartheta_{n}(\vec{p},\vec{g},t)|g\rangle \otimes |n+1\rangle\otimes |\vec{p}\rangle,
\end{equation}
\begin{equation}
|-,n,\vec{p}\rangle=\sin\vartheta_{n}(\vec{p},\vec{g},t)|e\rangle\otimes|n\rangle \otimes |\vec{p}\rangle
-\cos\vartheta_{n}(\vec{p},\vec{g},t)|g\rangle \otimes |n+1\rangle\otimes |\vec{p}\rangle,
\end{equation}
where
\begin{equation}
\sin\vartheta_{n}(\vec{p},\vec{g},t)\equiv\frac{\triangle(\vec{p},\vec{g},t)+\Omega_{n}(\vec{p},\vec{g},t)}{\sqrt{(\triangle(\vec{p},\vec{g},t)+\Omega_{n}(\vec{p},\vec{g},t))^{2}+4\lambda^{2}(n+1)}},
\end{equation}
\begin{equation}
\cos\vartheta_{n}(\vec{p},\vec{g},t)\equiv\frac{2\lambda\sqrt{n+1}}{\sqrt{(\triangle(\vec{p},\vec{g},t)+\Omega_{n}(\vec{p},\vec{g},t))^{2}+4\lambda^{2}(n+1)}},
\end{equation}
with
\begin{equation}
\Omega_{n}(\vec{p},\vec{g},t)\equiv\sqrt{\triangle(\vec{p},\vec{g},t)^{2}+4\lambda^{2}(n+1)},
\end{equation}
as the Rabi frequency. In Fig.1, the energy
eigenvalues is plotted as a
 function of the scaled time $\lambda t$ for three values of the
 parameter $\vec{q}.\vec{g}$. In this figure and all the
 subsequent figures we set
$q=10^{7}m^{-1}$, $M=10^{-26}Kg$, $g=9.8\frac{m}{s^{2}}$,
$\omega_{rec}=\frac{\hbar q^{2}}{2M}=.5\times10^{6}\frac{rad}{s}$,
$\lambda=1\times 10^{6}\frac{rad}{s}$ and $\omega_{c}=9\times10^{7}\frac{rad}{s}$ [24-27]. In
Figs.1a, 1b and 1c we consider the gravitational influence for
$\vec{q}.\vec{g}=0.5\times 10^{7}$, $\vec{q}.\vec{g}=1 \times 10^{7}$ and $\vec{q}.\vec{g}=1.5\times
10^{7}$. By comparing Figs.1a, 1b and 1c we can see the
influence of gravity on the time evolution of the energy
eigenvalues. With the increasing value of the parameter
$\vec{q}.\vec{g}$, the energy eigenvalues separation decrease.\\
\hspace*{00.5 cm} Now, we consider the effective mass $m^{*}$ for
  a moving two-level atom of mass $M$
  exposed simultaneously to a single-mode
   traveling wave field and a homogeneous gravitational field [5]
\begin{equation}
m^{*}=|\frac{\partial^{2}E_{\pm,n} }{\partial p ^{2} }  |^{-1}.
\end{equation}
 By using equations (6) and (12) the effective mass of the system is obtained
\begin{equation}
m^{*}(\vec{p},\vec{g},t)=\eta^{-3}\Omega_{n}^{3}(\vec{p},\vec{g},t),
\end{equation}
where $\eta=(\frac{4\lambda^{2}\hbar(n+1)q^{2}\cos^{2}\theta
}{M^{2}} )^{\frac{1}{3} } $, and $\theta$ is the angle between
$\vec{q}$ and $\vec{p}$. The time evolution of the
effective mass of the system has been shown in Figs.2a, 2b and 2c for
three values of the parameter $\vec{q}.\vec{g}$. With the increasing value of the parameter
$\vec{q}.\vec{g}$, the effective mass of the system decreases.
\section{Dynamical Evolution and the Wigner Distribution of a Field Mode}
In section 2, the effective mass with an effective
Hamiltonian for the atom-field system in the presence of a
homogeneous gravitational field is obtained. In this section, dynamical evolution of the system will be investigated.
How the gravitational field may affect the Wigner distribution of a field mode will be showed. For this purpose,
the Schr\"{o}dinger equation
\begin{equation}
i \hbar \frac{\partial |\psi \rangle } {\partial
 t}=\hat{\tilde{H}}_{eff}|\psi\rangle ,
\end{equation}
for the state vector $|\psi(t)\rangle$ with the Hamiltonian (2) will be solved.
At time $t=0$ the atom is uncorrelated with the field  and the
state vector of the system can be written as a direct product
\begin{eqnarray}
|\psi(t=0)\rangle = &&|\psi_{c.m}(0)\rangle
\otimes|\psi_{atom}(0)\rangle \otimes|\psi_{field}(0)\rangle
\nonumber
\\=&& (\int d^{3}p
\phi(\vec{p})|\vec{p}\rangle)\otimes(c_{e}|e\rangle
+c_{g}|g\rangle )\otimes(\sum_{n=0}w_{n}|n\rangle),
\end{eqnarray}
where we have assumed that initially the field is in a coherent
superposition of Fock states, the atom is in a coherent
superposition of its excited and ground states, and the wave
vector for the center-of-mass degree of freedom is
$|\psi_{c.m}(0)\rangle=\int d^{3}p \phi(\vec{p})|\vec{p}\rangle$.
In notation (15) we have
\begin{eqnarray}
|\psi(t=0)\rangle=&&\int d^{3}p
\sum_{n=0}(\psi_{1,n}(\vec{p},\vec{g},t=0)|e\rangle \otimes|n\rangle \otimes
|\vec{p}\rangle \nonumber \\+&&
\psi_{2,n}(\vec{p},\vec{g},t=0)|g\rangle\otimes |n\rangle \otimes|\vec{p}\rangle),
\end{eqnarray}
where the initial conditions are found
\begin{equation}
\psi_{1,n}(\vec{p},\vec{g},t=0)=w_{n}c_{e}\phi(\vec{p})
,\psi_{2,n}(\vec{p},\vec{g},t=0)=w_{n}c_{g}\phi(\vec{p}).
\end{equation}
At any time $t>0$, the atom-field state in the presence of a
homogeneous gravitational field is described by the state
\begin{equation}
|\psi(t)\rangle=\int d^{3}p
\sum_{n=0}(\psi_{1,n}(\vec{p},\vec{g},t)|e\rangle \otimes|n\rangle \otimes
|\vec{p}\rangle + \psi_{2,n}(\vec{p},\vec{g},t)|g\rangle\otimes |n\rangle
\otimes|\vec{p}\rangle).
\end{equation}
By solving the Schr\"{o}dinger equation (14) and by using equations (6), (7) and (8) we obtain
\begin{equation}
\psi_{1,n}(\vec{p},\vec{g},t)=\phi_{1,n}(\vec{p},\vec{g},t)\cos\vartheta_{n}(\vec{p},\vec{g},t)
+\phi_{2,n}(\vec{p},\vec{g},t)\sin\vartheta_{n}(\vec{p},\vec{g},t),
\end{equation}
\begin{equation}
\psi_{2,n}(\vec{p},\vec{g},t)=(\phi_{1,n-1}(\vec{p},\vec{g},t)\sin\vartheta_{n-1}(\vec{p},\vec{g},t)
-\phi_{2,n-1}(\vec{p},\vec{g},t)\cos\vartheta_{n-1}(\vec{p},\vec{g},t)),
\end{equation}
with
\begin{eqnarray}
\phi_{1,n}(\vec{p},\vec{g},t)=&&(\phi_{1,n}(\vec{p},\vec{g},t=0)A_{0}(\vec{p},\vec{g},t=0) exp(-i\omega_{c}(n+1)t)\nonumber \\\times&&
exp(i[\frac{\triangle(\vec{p},\vec{g},t)\Omega_{n}(\vec{p},\vec{g},t)}{2\vec{q}.\vec{g}}
\nonumber \\ +&& \frac{2\lambda^{2}(n+1)}{\vec{q}.\vec{g}}log(\triangle(\vec{p},\vec{g},t)+\Omega_{n}(\vec{p},\vec{g},t)]),
\end{eqnarray}
\begin{eqnarray}
\phi_{2,n}(\vec{p},\vec{g},t)=&&(\phi_{2,n}(\vec{p},\vec{g},t=0)A_{0}^{*}(\vec{p},\vec{g},t=0) exp(-i\omega_{c}(n+1)t)\nonumber \\ \times &&
exp(-i[\frac{\triangle(\vec{p},\vec{g},t)\Omega_{n}(\vec{p},\vec{g},t)}{2\vec{q}.\vec{g}}
\nonumber \\ +&& \frac{2\lambda^{2}(n+1)}{\vec{q}.\vec{g}}log(\triangle(\vec{p},\vec{g},t)+\Omega_{n}(\vec{p},\vec{g},t)]),
\end{eqnarray}
where
\begin{eqnarray}
A_{0}(\vec{p},\vec{g},t=0)=&&
exp(-i[\frac{\triangle(\vec{p},\vec{g},t=0)\Omega_{n}(\vec{p},\vec{g},t=0)}{2\vec{q}.\vec{g}}
\nonumber \\ +&& \frac{2\lambda^{2}(n+1)}{\vec{q}.\vec{g}}log(\triangle(\vec{p},\vec{g},t=0)+\Omega_{n}(\vec{p},\vec{g},t=0)]),
\end{eqnarray}
and
\begin{eqnarray}
\phi_{1,n}(\vec{p},\vec{g},t=0)=&&\psi_{1,n}(\vec{p},\vec{g},t=0)\cos\vartheta_{n}(\vec{p},\vec{g},t=0)
\nonumber \\ +&&\psi_{2,n+1}(\vec{p},\vec{g},t=0)\sin\vartheta_{n}(\vec{p},\vec{g},t=0),
\end{eqnarray}
\begin{eqnarray}
\phi_{2,n}(\vec{p},\vec{g},t=0)=&&\psi_{2,n+1}(\vec{p},\vec{g},t=0)\cos\vartheta_{n}(\vec{p},\vec{g},t=0)
\nonumber \\ -&&\psi_{1,n}(\vec{p},\vec{g},t=0)\sin\vartheta_{n}(\vec{p},\vec{g},t=0),
\end{eqnarray}
where the initial conditions $\psi_{1,n}(\vec{p},\vec{g},t=0)$ and
$\psi_{2,n}(\vec{p},\vec{g},t=0)$ in (17) are defined.
Also, $\sin\vartheta_{n}(\vec{p},\vec{g},t=0)$ and $\cos\vartheta_{n}(\vec{p},\vec{g},t=0)$ in (9) and (10) are defined,
respectively.\\
 \hspace*{00.5 cm}Now, the Influence of a classical homogeneous gravitational field
   on the Wigner distribution of
the radiation field will be studied. Brune et al. [29]
 show that, at the atomic inversion half-revival time, due to the quantum interaction between the atom and the
    field, the Wigner distribution of the
     field mode has two positive blobs symmetrically with respect to the origin and one negative blob in the
     origin. Kenfack and Zyczkowski [30] show that,
     negativity of the Wigner function is an indicator of
     nonclassicality. Also, $t=t_{R}/2=7\pi/2\lambda$ [31] is found
     which corresponds to one-half of the revival time of the atomic inversion when the gravitational field
 is not taken into account. However, in these theoretical studies
     results are obtained only under the condition that the atomic motion is neglected and the influence
     of the gravitational field is not taken into account. The Wigner distribution is particularly convenient for displaying
simultaneously the energy and the phase information of a single
mode field in a very simple and graphic form. It also allows for
a simple analysis of the field. The Wigner distribution in the presence of the gravitational field
$W(\beta,\beta^{*},\vec{g},t )$ of the complex amplitude $\beta=X+iY$ is
the real two-dimensional Fourier transformation of the symmetric
characteristic function, defined as [32]
\begin{equation}
W(\beta,\beta^{*},\vec{g},t )=\frac{1}{\pi^{2} }\int\exp(-\frac{\gamma^{2}}{2})C_{N}(\gamma,\gamma^{*},\vec{g},t)\exp(\beta \gamma^{*}-\beta^{*}\gamma)d^{2}\gamma,
\end{equation}
where the field normally ordered characteristic function is given by
\begin{equation}
C_{N}(\gamma,\gamma^{*},\vec{g},t)=\langle\exp(\gamma \hat{a}^{\dagger})\exp(-\gamma^{*}\hat{a})\rangle=Tr_{f}[\hat{\rho}_{f}(\vec{g},t)\exp(\gamma \hat{a}^{\dag})\exp(-\gamma^{*} \hat{a})],
\end{equation}
where $\gamma$ is a c-number variable and at any time $t>0$, the reduced density operator of
the cavity-field is given by
\begin{eqnarray}
\hat{\rho} _{f}(\vec{g},t)=&&Tr_{a}[\hat{\rho} _{a-f}(\vec{g},t)]=\int d^{3}p
\langle\vec{p} |\otimes(\sum_{i=e,g}\langle i|\psi(t)\rangle \langle\psi(t)|i\rangle )\otimes|\vec{p}\rangle
 \nonumber \\=&&\int d^{3}p \sum_{n,m=0}((\phi_{1,n}(\vec{p},\vec{g},t)\cos\vartheta_{n}(\vec{p},\vec{g},t)
+\phi_{2,n}(\vec{p},\vec{g},t)\sin\vartheta_{n}(\vec{p},\vec{g},t))
 \nonumber \\ \times &&(\phi_{1,m}^{*} (\vec{p},\vec{g},t)\cos\vartheta_{m}(\vec{p},\vec{g},t)
+\phi_{2,m}^{*} (\vec{p},\vec{g},t)\sin\vartheta_{m}(\vec{p},\vec{g},t))
 \nonumber \\+&&((\phi_{1,n-1}(\vec{p},\vec{g},t)\sin\vartheta_{n-1}(\vec{p},\vec{g},t)
-\phi_{2,n-1}(\vec{p},\vec{g},t)\cos\vartheta_{n-1}(\vec{p},\vec{g},t)))
 \nonumber \\ \times &&((\phi_{1,m-1}^{*} (\vec{p},\vec{g},t)\sin\vartheta_{m-1}(\vec{p},\vec{g},t)
-\phi_{2,m-1}^{*} (\vec{p},\vec{g},t)
\nonumber \\ \times &&\cos\vartheta_{m-1}(\vec{p},\vec{g},t))))|n\rangle \langle m|,
\end{eqnarray}
where $\sin\vartheta_{n}(\vec{p},\vec{g},t)$,
$\cos\vartheta_{n}(\vec{p},\vec{g},t)$, $\phi_{1,n}(\vec{p},\vec{g},t)$ and $\phi_{2,n}(\vec{p},\vec{g},t)$
in (9), (10), (21) and (22) are defined, respectively.
 We assume at $t=0$, the two-level atom is in a coherent superposition of the excited state and the
ground state with $c_{g}(0)=\frac{1}{\sqrt{2}}$,
$c_{e}(0)=\frac{1}{\sqrt{2}}$. We now consider the influence of a classical homogeneous gravitational field
on the Wigner distribution of
the radiation field when at $t=0$, the
cavity-field is prepared in a Glauber coherent state,
$w_{n}(0)=\frac{\exp(-\frac{|\alpha|^{2}}{2})\alpha^{n}}{\sqrt{n!}}$.
 In Fig.3 the Wigner distribution of
the radiation field as a
 function of the scaled time $\lambda t$ is plotted. We consider $\phi(\vec{p})=\frac{1}{\sqrt{2\pi
\sigma_{0}}}\exp(\frac{-p^{2}}{\sigma_{0}^{2}})$ with
$\sigma_{0}=1$, $\alpha=5$ [27] and $t=t_{R}/2=7\pi/2\lambda$ [31].
 In Fig.3a the Wigner distribution of
the radiation field with
small gravitational influence will be considered. This means very small
$\vec{q}.\vec{g}$, i.e., the momentum transfer from the laser
beam to the atom is only slightly altered by the gravitational
acceleration because the latter is very small or nearly
perpendicular to the laser beam. In Fig.3a,
 the Wigner distribution has both negative and positive values for $\vec{q}.\vec{g}=0$.
In Figs.3b and 3c the evolution of the Wigner distribution of
the radiation field for $\vec{q}.\vec{g}=0.5\times 10^{7}$ and
  $\vec{q}.\vec{g}=1.5\times 10^{7}$ will be considered, respectively. In Figs.3b and 3c, with the increasing value of the parameter
  $\vec{q}.\vec{g}$, the nonclassical behavior of the Wigner distribution of the cavity-field is suppressed.\\
\\
\section{Summary and conclusions}
The effective mass that approximately describes the effect of a classical homogeneous gravitational field on an
interacting atom-radiation field system determined within the framework of the Jaynes-Cummings model.
 By taking into account both the atomic motion and gravitational field, a full quantum treatment
 of the internal and external dynamics of the atom presented. By solving exactly the Schr\"{o}dinger
 equation in the interaction picture, the evolving state of the system found.
  Influence of a classical homogeneous gravitational field on the energy eigenvalues, the effective mass of
   atom-radiation field system
    and the Wigner distribution of the radiation field studied, when initially
  the radiation field is prepared in a coherent state and the two-level atom is
in a coherent superposition of the excited and ground states.
The results are summarized as follows: With the increasing value of the parameter
$\vec{q}.\vec{g}$, 1) the energy eigenvalues separation decrease 2) the effective mass of the
system decreases and 3) the nonclassical behavior of the Wigner distribution of the cavity-field is suppressed.\\
\\
{\bf  Acknowledgements} \\
The author wishes to thank The Office of Research
of the Islamic Azad University - Shahreza Branch for their
support.

\vspace{20cm}

{\bf FIGURE CAPTIONS:}

{\bf FIG. 1 } The time evolution of the energy
eigenvalues versus the scaled time $\lambda t$. Here we have set
$q=10^{7}m^{-1}$,\\
$M=10^{-26}kg$,$g=9.8\frac{m}{s^{2}}$,$\omega_{rec}=.5\times10^{6}\frac{rad}{s}$,\\$\lambda=1\times
10^{6}\frac{rad}{s}$,
$\omega_{c}=9\times10^{7}\frac{rad}{s}$,
 and $\alpha=5$;

 {\bf a)}For $\vec{q}.\vec{g}=0.5 \times 10^{7}$.

{\bf b)}For $\vec{q}.\vec{g}=1 \times 10^{7}$.

{\bf c)}For $\vec{q}.\vec{g}=1.5 \times 10^{7}$.\\

{\bf FIG. 2 }The time evolution of the effective mass of
atom-radiation field
 system versus the scaled time $\lambda t$;\\

{\bf a)}For $\vec{q}.\vec{g}=0.5 \times 10^{7}$.

{\bf b)}For $\vec{q}.\vec{g}=1 \times 10^{7}$.

{\bf c)}For $\vec{q}.\vec{g}=1.5 \times 10^{7}$.\\

{\bf FIG. 3 } The Wigner distribution of the cavity-field versus
$X=Re(\beta)$ and $Y=Im(\beta)$  with the same corresponding data
 used in Fig.1 and $t={\frac{t_{R}}{2}}=\frac{7\pi}{2\lambda}$;\\

{\bf a)}For $\vec{q}.\vec{g}=0$.

{\bf b)}For $\vec{q}.\vec{g}=1 \times 10^{7}$.

{\bf c)}For $\vec{q}.\vec{g}=1.5 \times 10^{7}$.\\

\end{document}